\documentclass[superscriptaddress,showpacs,a4paper]{revtex4}
\usepackage{subeqn}
\usepackage{graphicx}
\usepackage{numprint}
\usepackage[latin1]{inputenc}

\def\br{{\bf r}}

\newcommand{\EVRY}{Universit\'e d'Evry-Val d'Essonne, Boulevard François Mitterrand, 91025 Evry Cedex, France}
\newcommand{\LKB}{Laboratoire Kastler Brossel, UPMC-Paris 6, ENS, CNRS ; Case 74, 4 place Jussieu, 75005 Paris, France}
\newcommand{\bN}{\mbox{\boldmath$\nabla$}}
\npdecimalsign{.}

\begin{document}
\title{Analytical matrix elements of the Uehling potential in three-body systems, and applications to exotic molecules}

\author{Jean-Philippe Karr}
\email{karr@spectro.jussieu.fr}
\affiliation{\LKB}
\affiliation{\EVRY}

\author{Laurent Hilico}
\affiliation{\LKB}
\affiliation{\EVRY}

\date{\today}
\begin{abstract}
Exact analytical expressions for the matrix elements of the Uehling potential in a basis of explicitly correlated exponential wave functions are presented. The obtained formulas are then used to compute with an improved accuracy the vacuum polarization correction to the binding energy of muonic and pionic molecules, both in a first-order perturbative treatment and in a nonperturbative approach. The first resonant states lying below the $n=2$ threshold are also studied, by means of the stabilization method with a real dilatation parameter.
\end{abstract}
\maketitle

\section{Introduction}

The involvement of muonic molecular ions in nuclear fusion as fusion catalysts, through the Vesman mechanism~\cite{vesman1967}, generated great interest for precise energy level calculations in small muonic molecules~\cite{ponomarev2001}. In particular, precise knowledge of the binding energy of the weakly bound state $(L=1,v=1)$ in $dd\mu$ and $dt\mu$ is required to predict the temperature dependance of molecular formation rates. The analysis of $dd\mu$ fusion experiments performed at PNPI actually resulted in a very precise determination of the $(L=1,v=1)$ binding energy (with 0.7 meV uncertainty), in impressive agreement with theory~\cite{balin2011}. Knowledge of the spectrum of resonant states below the $n=2$ threshold is also useful for evaluating their impact in the muon catalyzed fusion cycle~\cite{froelich1995,wallenius1996}.

Exotic molecular ions also play a role in the interpretation of spectroscopy experiments in muonic or pionic atoms. The existence of $\mu p$ atoms in the metastable $2S$ state, a prerequisite for the measurement of the $2S$-$2P$ Lamb shift~\cite{pohl2010} was observed through a quenching mechanism by collisions with $H_2$ which involves resonant states of $pp\mu$ below the $n=2$ threshold~\cite{pohl2006}. In experiments on pionic hydrogen or deuterium~\cite{gotta2012}, atoms are produced from highly excited states through an atomic cascade in which resonances of $pp\pi$ or $dd\pi$ may be populated~\cite{jonsell1999}; the properties of these resonances are useful input parameters for an accurate modeling of the atomic cascade, which is indispensable to understand the observed line shape and extract the strong interaction broadening.

Some of these applications (most notably muon catalyzed fusion studies) require accurate energy level calculations, which means that leading corrections to the nonrelativistic energies have to be taken into account. In muonic systems, by far the largest correction originates from the vacuum polarization contribution to the interaction energy, whereas in pionic systems the strong interaction shift is of the same order~\cite{jonsell1999}. The first-order polarization correction to the interaction potential is usually referred to as the Uehling potential~\cite{uehling1935}; it is given by a nonelementary integral over a parameter. Most calculations of the Uehling correction in three-body systems have been performed by means of a numerical integration of its matrix elements, either with a Gaussian~\cite{wallenius1996,jonsell1999,myint1989} or exponential~\cite{aissing1990a,frolov2011} basis set. An analytical expression of its matrix elements in a correlated exponential basis set was published in Ref.~\cite{petelenz1987}. However, that expression is quite complicated, and numerical results obtained from it~\cite{petelenz1989} are in disagreement with those of other authors.

In the present work, we give in Sec.~\ref{deriv-uehling} a more compact analytic expression for the matrix elements of the Uehling potential in a correlated exponential basis set, which greatly simplifies its application in actual calculations. These results may also be applied to calculations with the generalized Hylleraas expansion~\cite{drake}. The obtained expressions are then used in Sec.~\ref{num} to obtain a new set of reference results for bound and resonant state energies in muonic and pionic molecules.

\section{Matrix elements of the Uehling potential} \label{deriv-uehling}

We use atomic units, scaled to the mass $m$ of the lightest particle of the studied three-body system (e.g. the muon mass in the case of muonic molecules). The Uehling potential between two particles of charges $Z_1,Z_2$ reads~\cite{uehling1935}
\begin{equation}
V_{vp}(r) = \frac{\alpha_{fsc} Z_1 Z_2}{3\pi r} \int_1^{\infty} du \, e^{-2xru} \frac{\sqrt{u^2 - 1} \left( 2u^2 + 1 \right)}{u^4} \label{expr-uehling}
\end{equation}
with $x = \left( \alpha_{fsc} m \right)^{-1}$ (here $\alpha_{fsc}$ represents the fine structure constant). We consider a variational expansion of the three-body wavefunction in the form
\begin{equation}
\Psi \left( \br_1,\br_2,\br_{12} \right) = \sum_{n=1}^{N} C_n \; e^{- \alpha_n r_1 - \beta_n r_2 - \gamma_n r_{12}} \; \mathcal{Y}^{l_1l_2}_{LM}(\hat\br_1,\hat\br_2), \label{expansion}
\end{equation}
where $r_1$, $r_2$, $r_{12}$ are the interparticle distances and $\mathcal{Y}^{l_1l_2}_{LM}(\hat\br_1,\hat\br_2)$ are bipolar spherical harmonics. $\alpha_n$, $\beta_n$, $\gamma_n$ are real exponents satisfying the relations $\alpha_n + \beta_n > 0$, $\alpha_n + \gamma_n > 0$, and $\beta_n + \gamma_n > 0$. The matrix elements of the Uehling potential in such a basis set involve integrals of the form
%
%
\begin{equation}
I_{l,m,n}^{(i)} (\alpha,\beta,\gamma) = \int\!\int\!\int dr_1 dr_2 dr_{12} \; r_1^l r_2^m r_{12}^n \; e^{- \alpha r_1 - \beta r_2 - \gamma r_{12}} \; \int_1^{\infty} du \, e^{-2xur_i} \frac{\sqrt{u^2 - 1} \left( 2u^2 + 1 \right)}{u^4}
\end{equation}
where $r_i=r_1$, $r_2$, and $r_{12}$ for $V_{vp} \left( r_1 \right)$, $V_{vp} \left( r_2 \right)$ and $V_{vp} \left( r_{12} \right)$ respectively, and $l,m,n$ are non-negative integers. These integrals can be generated from $I_{0,0,0} (\alpha,\beta,\gamma)$ by partial differentiation with respect to $\alpha$, $\beta$, and $\gamma$, as it is usually done in the case of the Coulomb potential (see e.g.~\cite{sack1967,korobov2002}). The basic integral to be calculated is thus
\begin{equation}
I_{0,0,0}^{(i)} (\alpha,\beta,\gamma) = \int\!\int\!\int dr_1 dr_2 dr_{12} \; e^{- \alpha r_1 - \beta r_2 - \gamma r_{12}} \; \int_1^{\infty} du \, e^{-2xur_i} \frac{\sqrt{u^2 - 1} \left( 2u^2 + 1 \right)}{u^4}
\end{equation}
The first step is to change the order in which the integrations over space coordinates and the parameter $u$ are performed. For $V_{vp} \left( r_1 \right)$ one obtains
\begin{equation}
I_{0,0,0}^{(1)} (\alpha,\beta,\gamma) = \int_1^{\infty} du \frac{\sqrt{u^2 - 1} \left( 2u^2 + 1 \right)}{u^4} \; \int\!\int\!\int dr_1 dr_2 dr_{12} \; e^{- (\alpha+2xu) r_1 - \beta r_2 - \gamma r_{12}}
\end{equation}
The integral over spatial coordinates is well known~\cite{sack1967,korobov2002} and reads $2/(\beta + \gamma)(\alpha+\beta+2xu)(\alpha+\gamma+2xu)$, so that
\begin{equation}
I_{0,0,0}^{(1)} (\alpha,\beta,\gamma) = \frac{1}{2(\beta+\gamma)x^2} I_1(a,b) \; ,
\end{equation}
where $a = (\alpha+\beta)/2x$, $b = (\alpha+\gamma)/2x$, and
\begin{equation} \label{integral}
I_1(a,b) = \int_1^{\infty} du \frac{\sqrt{u^2-1} \left(2u^2 + 1 \right)}{u^4(u+a)(u+b)} \; .
\end{equation}
The integral~(\ref{integral}) can be obtained analytically by standard procedures (the work can be done using a symbolic computation program such as Mathematica):
\begin{eqnarray}
I_1(a,b) &=& \frac{3\pi (a+b) \left[ 2\left( a^2+b^2 \right) + 3a^2b^2 \right] - ab \left[ 12 \left( a^2 + ab + b^2 \right) + 20 a^2 b^2 \right]}{12 a^4 b^4} \nonumber \\
&+& \frac{\sqrt{1-a^2} \left( 1 + 2a^2 \right) \arccos(a)}{a^4 (a-b)} - \frac{\sqrt{1-b^2} \left( 1 + 2b^2 \right) \arccos(b)}{b^4 (a-b)} \; .
\end{eqnarray}
Since the last two terms in this expression diverge for $a=b$, one should study the limit $b \rightarrow a$. The result is
\begin{equation}
I_1(a,a) = \frac{3\pi \left( 4 + 3 a^2 \right) - 2a \left( 12 + 11 a^2 \right)}{6 a^5} + \frac{\left( 2a^4 - a^2 - 4 \right) \arccos(a)}{a^5 \sqrt{1-a^2}} \; .
\end{equation}
For $S$ states, the matrix elements of $V_{vp} \left( r_1 \right)$ involve the integral
\begin{equation}
I_{0,1,1}^{(1)} (\alpha,\beta,\gamma) = \frac{\partial^2 I_{0,0,0}^{(1)} (\alpha,\beta,\gamma)} {\partial \beta \, \partial \gamma}
\end{equation}
Straightforward (but tedious) algebraic manipulations lead to
\begin{equation}
I_{0,1,1}^{(1)} (\alpha,\beta,\gamma) = \frac{1} {(\beta+\gamma)x^2} \left[ \frac{I_1(a,b)}{(\beta+\gamma)^2} + \frac{I_2(a,b)}{4x(\beta + \gamma)} + \frac{I_3(a,b)}{8x^2} \right] \, ,
\end{equation}
where
\begin{eqnarray}
I_2(a,b) &=& \frac{3\pi \left[ 4\left( a^4+a^3b+a^2 b^2+ab^3+b^4 \right) + 3\left( a^4 b^2 + a^3 b^3 + a^2 b^4 \right) \right] - 2ab (a+b) \left[ 12 \left( a^2 + b^2 \right) + 11 a^2 b^2 \right]}{6 a^5 b^5} \nonumber \\
&+& \frac{ \left( 4 + a^2 -2a^4 \right) \arccos(a)}{a^5 (a-b) \sqrt{1-a^2}} - \frac{ \left( 4 + b^2 -2b^4 \right) \arccos(b)}{b^5 (a-b) \sqrt{1-b^2}}
\end{eqnarray}
\begin{eqnarray}
I_3(a,b) &=& \frac{3\pi\! (\!a\!+\!b\!)\! \left[ \!4\!\left(\! a^2\!+\!b^2\! \right)\! +\! 2 ab\! + \! 3 a^2 b^2 \!\right]\! -\! 2ab\! \left[\! 12\! \left(\! a^4\! +\! b^4\! \right)\! +\! 11\! \left(\! a^4\! b^2\! +\! a^2 \! b^4 \! \right)\! - \! 6 \! \left( \! a^3 b \!+\! a^2 b^2 \!+\! a b^3 \! \right) \!-\! 10 a^3 b^3 \! \right]\!/\!(a-b)^2}{6 a^5 b^5} \nonumber \\
&+& \frac{ \left( 4b - 6a + a^2 b - 3 a^3 -2 a^4 b + 6 a^5 \right) \arccos(a)}{a^5 (a-b)^3 \sqrt{1-a^2}} - \frac{ \left( 4a - 6b + a b^2 - 3 b^3 -2 a b^4 + 6 b^5 \right) \arccos(b)}{b^5 (a-b)^3 \sqrt{1-b^2}} \, .
\end{eqnarray}
In the case $a=b$, these expressions are replaced by
\begin{subequations}
\begin{equation}
I_2(a,a) = \frac{3\pi \left( 20 - 11 a^2 - 9 a^4 \right) - 2a \left( 60 - 23 a^2 - 28 a^4 \right)}{6 a^6 \left(1 - a^2\right) } - \frac{\left( 20 - 21 a^2 - 6 a^4 + 4 a^6 \right) \arccos(a)}{a^6 \left( 1-a^2 \right)^{3/2}}
\end{equation}
\begin{equation}
I_3(a,a) = \frac{3\pi \! \left( \! 20 \!+\! 6 \! a^2 \! \right) \! \left( \! 1 \! - \! a^2 \! \right)^2 \! - \! a \! \left( \!  120 \!-\! 184 a^2 \! +\! 23 a^4 \!+\! 32 a^6 \!\right)}{6 a^7 \left(1 - a^2\right)^2 } - \frac{\left(\! 40 \!-\! 88 a^2 \!+\! 45 a^4 \!+\! 10 a^6 \!-\! 4 a^8 \!\right) \! \arccos(a)}{2 a^7 \left( 1-a^2 \right)^{5/2}} \, .
\end{equation}
\end{subequations}
The matrix elements of $V_{vp} \left( r_2 \right)$ (resp. $V_{vp} \left( r_{12} \right)$) can be deduced from this result by interchange of the parameters $\alpha$ and $\beta$ (resp. $\alpha$ and $\gamma$).

For $P$ states, three integrals are needed: $I_{2,1,1}^{(1)}$, $I_{0,3,1}^{(1)}$, and $I_{0,1,3}^{(1)}$. Their expressions are too lengthy to be reported here, but they can be easily evaluated by symbolic calculations programs and translated into C or FORTRAN code. For higher values of the orbital angular momentum, it is doubtful whether evaluation of analytical formulas remains advantageous with respect to numerical integration, because of growing calculation time and numerical instabilities.

\section{Numerical approach and results} \label{num}

\subsection{Numerical approach} \label{method}

In this Section, we present the results of variational calculations using the non relativistic three-body Hamiltonian
\begin{equation}
H = -\frac{1}{2m_{1}}\!\bN^2_{\br_1}\!-\!\frac{1}{2m_{2}}\!\bN^2_{\br_2}\!-\!\frac{1}{m}\!\bN_{\br_1}\!\bN_{\br_2}\! -\! \frac{1}{r_1}\!-\!\frac{1}{r_2}\!+\!\frac{1}{r_{12}}.
\end{equation}
Here, the nuclei are numbered by 1 and 2, and the light particle (muon or pion) by 3. The notations $r_1 \equiv r_{13}$, $r_2 \equiv r_{23}$ are used. $m_1$ and $m_2$ are respectively the $1-3$ and $2-3$ reduced masses. The vacuum polarization correction to the binding energy is determined from first-order perturbation theory:
\begin{equation}
\Delta E_b^{(1)} = \Delta E_{at}^{(1)} - \left( \left\langle V_{vp}(r_1) \right\rangle + \left\langle V_{vp}(r_2) \right\rangle + \left\langle V_{vp}(r_{12}) \right\rangle \right)
\end{equation}
where $\Delta E_{at}^{(1)}$ is the first-order shift of the related atomic threshold.

The Uehling potential behaves like $\ln(r)/r$ at $r \rightarrow 0$~\cite{blomqvist1972}. This not too singular behavior enables good convergence of a nonperturbative calculation, where the vacuum polarization potential $V_{vp}(r_1) + V_{vp}(r_2) + V_{vp}(r_{12}$) is directly added to the Coulomb Hamiltonian $H$ before diagonalization. The correction to the binding energy is then
\begin{equation}
\Delta E_b = E_{at}^{(CU)} - E^{(CU)} - (E_{at}^{(C)} - E^{(C)})
\end{equation}
where $E^{(C)(CU)}$ and $E_{at}^{(C)(CU)}$ are the energies of the three-body state and of the related atomic threshold, obtained with the Coulomb (C) or Coulomb+Uehling (CU) potential. While corrections beyond the first order are not useful in themselves at the present level of theoretical accuracy, this provides a simple and reliable way of evaluating higher-order corrections and thus controlling the accuracy of the results. In addition, the perturbative approach fails for weakly bound resonant states close to a $n \geq 2$ atomic threshold. One way to understand this is to consider that the lifting of the atomic manifold degeneracy induced by the Uehling potential modifies the long range behavior of the atom-nucleus interaction potential, from a $1/r^2$ dipole potential to a $1/r^4$ induced dipole potential. A nonperturbative calculation is thus mandatory in such cases~\cite{jonsell1999,wallenius1996b,karr2012}. In all the tables below, we give both the first-order perturbation result $\Delta E_b^{(1)}$ and higher-order corrections $\Delta E_b^{(>1)} = \Delta E_b - \Delta E_b^{(1)}$.

The expansion~(\ref{expansion}) was used, with real exponents $\alpha_n$, $\beta_n$ and $\gamma_n$ generated in a pseudorandom way in intervals $[A_1,A_2]$, $[B_1,B_2]$ and $[C_1,C_2]$ respectively~\cite{frolov1984,alexander1988}. Here the variational parameters are the bounds of the intervals, and were optimized separately for each calculated level. Basis sets of $N=1000-2500$ vectors were used to obtain good convergence of the results.

It should be noted that complex exponents are generally better suited for molecular systems~\cite{korobov2000}. The analytical formulas of Sec.~\ref{deriv-uehling} are still valid for complex exponents $\alpha_n$, $\beta_n$, $\gamma_n$, provided their real parts satisfy the relationships ${\rm Re}[\alpha_n + \beta_n] > 0$, ${\rm Re}[\alpha_n + \gamma_n] > 0$, and ${\rm Re}[\beta_n + \gamma_n] > 0$. However, with an expansion that uses complex exponents and/or using complex coordinate rotation~\cite{reinhardt1982} to study resonant states, numerical problems appear when the Uehling potential is included in the Hamiltonian. This suggests that the Uehling potential may not be dilation analytic~\cite{reed}. A rigorous analysis of this point is beyond the scope of the present paper, but would certainly be useful for further studies with the Uehling potential.

For nonperturbative calculations, it is important to add higher exponents in the basis set in order to describe accurately the behavior of the Uehling potential at small $r$. This is done by adding several subsets defined by
\begin{equation}
\left\{
\begin{array}{@{}l}
A_1^{(0)} = A_2,\\[1mm]
A_1^{(n)} = \tau^n A_1^{(0)},
\end{array}
\right.
\qquad
\left.
\begin{array}{@{}l}
A_2^{(0)} = \tau A_1^{(0)},\\[1mm]
A_2^{(n)} = \tau^n A_2^{(0)},
\end{array}
\right.
\end{equation}
Typically $\tau \sim 3-5$, and $n_{max} = 1-2$. We add similar basis sets for $r_2$ (if the basis is not symmetrized) and $r_{12}$.

With the above-mentioned typical basis size, quadruple-precision arithmetic is generally required to maintain sufficient numerical stability. However, the derived expressions of the Uehling potential's matrix elements are numerically unstable (for $a \approx b$), so that sextuple-precision arithmetic had to be used in most cases. For the weakly bound $(L=1,v=1)$ states in $dd\mu$ and $dt\mu$, which  require the largest basis sets, octuple precision proved necessary.

We used the latest CODATA (2010) values~\cite{codata} of the particle masses (muon, proton, deuteron and triton) and of the fine structure constant. For the pion mass, the latest value from the Particle Data Group~\cite{pdg}, was used. The quantity $x$ appearing in the expression~(\ref{expr-uehling}) of the Uehling potential is $x_{\mu} = 0.662 751 541 1$ for muonic systems, and $x_{\pi} = 0.501 720 701 5$ for pionic systems.

\subsection{Results}

We first determined the vacuum polarization shift of the $1S$ and $2S$ atomic thresholds, both in the perturbative and nonperturbative approaches, using a variational approach similar to the one described above. The radial atomic wavefunction $\Psi(r)$ is expanded on a set of $N=50-100$ exponentials $e^{-\alpha_n r}$ with pseudorandomly chosen real exponents. Results are summarized in Table~\ref{threshold}.
\begin{table}[h]
\begin{center}
\begin{tabular}{@{\hspace{5mm}}c@{\hspace{5mm}}c@{\hspace{5mm}}n{2}{7}@{\hspace{5mm}}n{2}{7}@{\hspace{5mm}}}
\hline\hline
\vrule width0pt height10pt depth4pt
Atom & state & \multicolumn{1}{c}{$\Delta E^{(1)}$} &  \multicolumn{1}{c}{$\Delta E$} \\
\hline
\vrule width0pt height10pt depth4pt
$\mu p$ & $1S$ & -1.898 829 6 & -1.900 865 8 \\
        & $2S$ & -0.219 584 0 & -0.219 737 2 \\
$\mu d$ & $1S$ & -2.129 272 6 & -2.131 642 2 \\
        & $2S$ & -0.245 319 4 & -0.245 494 5 \\
$\mu t$ & $1S$ & -2.214 430 5 & -2.216 926 1 \\
        & $2S$ & -0.254 804 0 & -0.254 987 2 \\
$\pi p$ & $1S$ & -3.240 801 9 & -3.244 916 5 \\
        & $2S$ & -0.368 276 3 & -0.368 560 0 \\
$\pi d$ & $1S$ & -3.732 175 0 & -3.737 120 2 \\
        & $2S$ & -0.422 196 4 & -0.422 529 5 \\
\hline\hline
\end{tabular}
\end{center}
\caption{Vacuum polarization shift of the $1S$ and $2S$ atomic states of muonic and pionic atoms, in eV. Both the first-order perturbation result $\Delta E^{(1)}$ and the nonperturbative result $\Delta E$ are given.}\label{threshold}
\end{table}

Table~\ref{muonic-bs} gives the energies of all the bound states of muonic molecules with orbital angular momentum $L=0,1$. The results are in perfect agreement with earlier calculations~\cite{aissing1990b}, with an accuracy improved from 0.1 meV to 1 $\mu$ eV. The contribution from higher perturbation orders is also obtained (for the first time to our knowledge), and typically amounts to a fraction of meV for the ground vibrational state. Precise experimental results are available only for the $(L=1,v=1)$ state of $dd\mu$~\cite{balin2011}, where there is good agreement with theoretical predictions~\cite{bakalov2001,harston1997,korobov2004} that also take leading relativistic and nuclear structure corrections, as well as corrections caused by the finite size of the $(dd\mu)dee$ molecular complex. The discrepancy is only 0.5 meV, while experimental and theoretical uncertainties are respectively of 0.7 and 0.4 meV. The 0.097 meV difference (-8.657 meV instead of -8.56 meV) between our new result for the Uehling correction, and the value of~\cite{bakalov2001} does not alter the agreement with experimental data. The newly obtained contribution from higher perturbation orders (0.003 meV) is currently not relevant in view of the overall theoretical uncertainty.
\begin{table}[h]
\begin{center}
\begin{tabular}{@{\hspace{5mm}}c@{\hspace{5mm}}c@{\hspace{5mm}}c@{\hspace{5mm}}n{3}{6}@{\hspace{5mm}}n{3}{3}@{\hspace{5mm}}n{1}{3}@{\hspace{5mm}}}
\hline\hline
\vrule width0pt height10pt depth4pt
Molecule & $L$ & $v$ & \multicolumn{1}{c}{$E_b$} & \multicolumn{1}{c}{$\Delta E_b^{(1)}$} & \multicolumn{1}{c}{$\Delta E_b^{(>1)}$} \\
         &     &     & \multicolumn{1}{c}{\scriptsize(eV)}  & \multicolumn{1}{c}{\scriptsize(meV)}              & \multicolumn{1}{c}{\scriptsize(meV)}               \\
\hline
\vrule width0pt height10pt depth4pt
$pp\mu$ & 0 & 0 & 253.150 104 & 284.875 & 0.430  \\
        & 1 & 0 & 107.265 303 &  50.581 & 0.089  \\
$pd\mu$ & 0 & 0 & 221.547 587 & 234.419 & 0.376  \\
        & 1 & 0 &  97.497 678 &  21.445 & 0.053  \\
$pt\mu$ & 0 & 0 & 213.838 459 & 222.385 & 0.365 \\
        & 1 & 0 &  99.126 024 &  21.009 & 0.055  \\
$dd\mu$ & 0 & 0 & 325.070 580 & 412.131 & 0.657  \\
        & 0 & 1 &  32.844 224 &  39.129 & 0.074  \\
        & 1 & 0 & 226.679 812 & 226.216 & 0.358  \\
        & 1 & 1 &   1.974 980 &  -8.657 & 0.003  \\
$dt\mu$ & 0 & 0 & 319.136 858 & 402.275 & 0.653 \\
        & 0 & 1 &  34.834 420 &  28.074 & 0.061  \\
        & 1 & 0 & 232.469 701 & 233.597 & 0.377  \\
        & 1 & 1 &   0.660 329 & -16.604 & 0.013  \\
$tt\mu$ & 0 & 0 & 362.906 436 & 480.211 & 0.781  \\
        & 0 & 1 &  83.770 686 &  99.858 & 0.172  \\
        & 1 & 0 & 107.265 303 & 331.988 & 0.534  \\
        & 1 & 1 &  45.205 712 &  34.072 & 0.072  \\
\hline\hline
\end{tabular}
\end{center}
\caption{Vacuum polarization correction to the binding energies for bound states of muonic molecules, obtained using the variational expansion~(\ref{expansion}) with real exponents. The binding energy $E_b$ calculated with the pure Coulomb potential is given in the fourth column. The vacuum polarization shift at first order of perturbation theory is given in the next column. The last column shows the difference between results of non-perturbative and first-order perturbative treatments. }\label{muonic-bs}
\end{table}

Results for the bound states of pionic molecules are given in Table~\ref{pionic-bs}. We have limited our study to $dd\pi$ and $pp\pi$, which could play a role in the interpretation of pionic hydrogen and deuterium spectroscopy experiments~\cite{gotta2012}. It should be noted that accuracy is much less essential than for muonic systems, because (i) experimental resolution is limited to about 10 $\mu$eV by the pion lifetime $\tau = 26$ ns, and (ii) theoretical accuracy is limited to a fraction of meV by the 2.5 ppm relative uncertainty on the pion mass. However, the vacuum polarization correction is relevant since it typically amounts to a fraction of eV for the ground vibrational state.

\begin{table}[h]
\begin{center}
\begin{tabular}{@{\hspace{5mm}}c@{\hspace{5mm}}c@{\hspace{5mm}}c@{\hspace{5mm}}n{3}{6}@{\hspace{5mm}}n{3}{3}@{\hspace{5mm}}n{1}{3}@{\hspace{5mm}}}
\hline\hline
\vrule width0pt height10pt depth4pt
Molecule & $L$ & $v$ & \multicolumn{1}{c}{$E_b$} & \multicolumn{1}{c}{$\Delta E_b^{(1)}$} & \multicolumn{1}{c}{$\Delta E_b^{(>1)}$} \\
         &     &     & \multicolumn{1}{c}{\scriptsize(eV)}  & \multicolumn{1}{c}{\scriptsize(meV)}              & \multicolumn{1}{c}{\scriptsize(meV)}               \\
\hline
\vrule width0pt height10pt depth4pt
$pp\pi$ & 0 & 0 & 294.859 450 & 431.020 & 0.763  \\
        & 1 & 0 &  80.227 512 &   6.808 & 0.055  \\
$dd\pi$ & 0 & 0 & 392.301 211 & 660.791 & 1.237  \\
        & 0 & 1 &  15.777 113 &  19.426 & 0.053  \\
        & 1 & 0 & 237.301 428 & 291.614 & 0.556  \\
\hline\hline
\end{tabular}
\end{center}
\caption{Same as Table~\ref{muonic-bs}, for bound states of the pionic molecules $pp\pi$ and $dd\pi$.}\label{pionic-bs}
\end{table}

In the following, we consider quasibound states (or resonances). In view of the problems with complex coordinate rotation mentioned in Sec.~\ref{method}, we used the stabilization technique with a real dilatation parameter, similarly to Ref.~\cite{wallenius1996b}. The accuracy of this method is limited by the width of the resonances. In the following tables, we report the widths calculated in Ref.~\cite{kilic2004} in order to explain the accuracy of the results. While a complete investigation of the resonance spectrum would lie beyond the scope of this paper, we give illustrative results for the first two vibrational and rotational states below the $2S$ threshold.

Among the muonic molecules, we have considered $dd\mu$ and $dt\mu$, in which fusion research has been the most active (see Table~\ref{muonic-res}). The involvement of resonances was originally proposed in the framework of $d-t$ fusion, whereas its impact in $d-d$ fusion is expected to be much less important~\cite{froelich1995,froelich1993}. In the case of $dt\mu$, our results are in good agreement with those of Ref.~\cite{wallenius1996b}, and represent an improvement in accuracy by 2-3 orders of magnitude.

\begin{table}[h]
\begin{center}
\begin{tabular}{@{\hspace{5mm}}c@{\hspace{5mm}}c@{\hspace{5mm}}c@{\hspace{5mm}}n{3}{6}@{\hspace{5mm}}n{2}{1}@{\hspace{5mm}}n{2}{3}@{\hspace{5mm}}n{2}{3}@{\hspace{5mm}}c@{\hspace{5mm}}}
\hline\hline
\vrule width0pt height10pt depth4pt
Molecule & $L$ & $v$ & \multicolumn{1}{c}{$E_b$} & \multicolumn{1}{c}{$\Gamma$}  & \multicolumn{1}{c}{$\Delta E_b^{(1)}$} & \multicolumn{1}{c}{$\Delta E_b$} & \multicolumn{1}{c}{$\Delta E_b$} \\
         &     &     & \multicolumn{1}{c}{\scriptsize(eV)} & \multicolumn{1}{c}{\scriptsize($\mu$eV)} & \multicolumn{1}{c}{\scriptsize(meV)} & \multicolumn{1}{c}{\scriptsize(meV)} & \multicolumn{1}{c}{\scriptsize(meV)} \\
         &     &     & & \multicolumn{1}{c}{\scriptsize \cite{kilic2004}} & & \multicolumn{1}{c}{\scriptsize This work} & \multicolumn{1}{c}{\scriptsize \cite{wallenius1996b}} \\
\hline
\vrule width0pt height10pt depth4pt
$dd\mu$ & 0 & 0 & 218.111 60  &  1.9 & -54.77  & -54.79  &  \\
        & 0 & 1 & 135.279 02  &  5.8 & -82.76  & -82.79  &  \\
        & 1 & 0 & 211.924 50  &  5.8 & -58.44  & -58.46  &  \\
        & 1 & 1 & 130.350 1   & 15.3 &         & -85.5   &  \\
$dt\mu$ & 0 & 0 & 217.889 86  &  3.0 & -59.83  & -59.86  &  -60 \\
        & 0 & 1 & 139.731 40  &  7.2 & -86.66  & -86.70  &  -85 \\
        & 1 & 0 & 212.545 744 &  0.5 & -63.006 & -63.030 &  -63 \\
        & 1 & 1 & 135.379 516 &  0.9 & -89.069 & -89.104 &  -91 \\
\hline\hline
\end{tabular}
\end{center}
\caption{Vacuum polarization correction to the binding energy for resonant states of the muonic molecules $dd\mu$ and $dt\mu$ below the $n=2$ threshold, obtained using the variational expansion~(\ref{expansion}) with real exponents. The binding energy $E_b$ obtained with the pure Coulomb potential is given in the fourth column. The fifth column contains the resonance widths taken from~\cite{kilic2004}, which give a measure of the precision of the results. The vacuum polarization shift, both at first order of perturbation theory and in a nonperturbative treatment, are given in the next two columns. The first-order result is given only in the cases where the precision is sufficient to evidence the difference with the nonperturbative result.}\label{muonic-res}
\end{table}

Table~\ref{pionic-res} summarizes results for the pionic molecules $pp\pi$ and $dd\pi$, where resonant states play a role in the deexcitation cascade of pionic atoms~\cite{gotta2012}. In the case of $pp\pi$, our results are in good agreement with those of Ref.~\cite{jonsell1999}, and bring an improvement in accuracy by 1-2 orders of magnitude. Note that the binding energies of the resonances we have studied are large enough for the perturbative approach to yield precise results. The difference with the result of a nonperturbative calculation is typically of 20-40~$\mu$eV only.

\begin{table}[h]
\begin{center}
\begin{tabular}{@{\hspace{5mm}}c@{\hspace{5mm}}c@{\hspace{5mm}}c@{\hspace{5mm}}n{3}{5}@{\hspace{5mm}}n{1}{4}@{\hspace{5mm}}n{3}{2}@{\hspace{5mm}}n{3}{2}@{\hspace{5mm}}n{3}{0}@{\hspace{5mm}}}
\hline\hline
\vrule width0pt height10pt depth4pt
Molecule & $L$ & $v$ & \multicolumn{1}{c}{$E_b$} & \multicolumn{1}{c}{$\Gamma$}  & \multicolumn{1}{c}{$\Delta E_b^{(1)}$} & \multicolumn{1}{c}{$\Delta E_b$} & \multicolumn{1}{c}{$\Delta E_b$} \\
         &     &     & \multicolumn{1}{c}{\scriptsize(eV)} & \multicolumn{1}{c}{\scriptsize($\mu$eV)} & \multicolumn{1}{c}{\scriptsize(meV)} & \multicolumn{1}{c}{\scriptsize(meV)} & \multicolumn{1}{c}{\scriptsize(meV)} \\
         &     &     & & \multicolumn{1}{c}{\scriptsize \cite{kilic2004}} & & \multicolumn{1}{c}{\scriptsize This work} & \multicolumn{1}{c}{\scriptsize \cite{jonsell1999}} \\
\hline
\vrule width0pt height10pt depth4pt
$pp\pi$ & 0 & 0 & 236.173     & 1.5    &         & -78     & -80  \\
        & 0 & 1 & 100.146     & 1.9    &         &  -136   & -140 \\
        & 1 & 0 & 220.381 8   & 0.20   &         &  -92.0  & -90  \\
        & 1 & 1 &  89.641     & 0.38   &         &  -145   & -150 \\
$dd\pi$ & 0 & 0 & 275.280 3   & 0.050  &         & -85.5   & \\
        & 0 & 1 & 156.821 8   & 0.097  &         & -139.7  & \\
        & 1 & 0 & 265.180 84  & 0.0041 & -93.87  & -93.90  & \\
        & 1 & 1 & 149.088 74  & 0.0054 & -145.56 & -145.60 & \\
\hline\hline
\end{tabular}
\end{center}
\caption{Same as Table~\ref{muonic-res}, for resonant states of the pionic molecules $pp\pi$ and $dd\pi$ below the $n=2$ threshold.}\label{pionic-res}
\end{table}

In conclusion, we have shown that the matrix elements of the Uehling potential in a basis of correlated exponential functions may be expressed in an analytical form. We have used the obtained expressions to calculate the vacuum polarization shift for a wide range of bound and resonant states in muonic and pionic molecules, either for the first time, or with a greatly improved accuracy. The excellent agreement with earlier calculations which used a numerical evaluation of matrix elements fully confirms the validity of the analytical formula.

\textbf{Acknowledgments.} We thank V.I. Korobov for sharing his program for variational calculations of three-body systems with exponential basis functions, and for helpful discussions.

\end{document}